\documentclass[11pt,a4paper]{article}

\usepackage[utf8]{inputenc}
\usepackage[T1]{fontenc}
\usepackage[english]{babel}
\usepackage{graphicx}
\usepackage{mathtools}
\usepackage{pos}
\usepackage{siunitx}
\usepackage{microtype}

\graphicspath{{./Graphics/}}
\renewcommand*{\hookAfterAffiliation}{\medskip}
\renewcommand*{\thefootnote}{\arabic{footnote}}
\newcommand{\+}{\hspace*{.08335em}}
\newcommand{\?}{\hspace*{-.08335em}}
\newcommand{\dd}{\mathrm{d}}
\newcommand{\qcond}{\langle\+\overline{\psi}\psi\rangle}
\newcommand{\sYM}{s_{\textup{gauge}}}
\newcommand{\muB}{\mu_{\textup{B}}}
\newcommand{\muBCEP}{\mu_{\textup{B}}^{\textup{CEP}}}
\newcommand{\TCEP}{T_{\textup{CEP}}}
\newcommand{\Tr}{\operatorname{Tr}}

\DeclareSIUnit{\MeV}{\mega\electronvolt}
\hypersetup{%
	pdftitle = {QCD's equation of state from Dyson-Schwinger equations},
	pdfauthor = {Philipp Isserstedt, Christian S. Fischer, Thorsten Steinert}
}

\AtBeginDocument{\setlength{\belowdisplayshortskip}{\belowdisplayskip}}

\title{QCD's equation of state from Dyson--Schwinger equations}

\author*[a,b,\dagger]{Philipp Isserstedt}
\author[\,a,b]{Christian S.~Fischer}
\author[\,a,\ddagger]{Thorsten Steinert}

\affiliation[a]{%
	Institut f\"{u}r Theoretische Physik,
	Justus-Liebig-Universit\"{a}t Gie\ss{}en,
	35392 Gie\ss{}en,
	Germany
}

\affiliation[b]{%
	Helmholtz Forschungsakademie Hessen f\"{u}r FAIR (HFHF),
	GSI Helmholtzzentrum f\"{u}r Schwerionenforschung,
	Campus Gie\ss{}en,
	35392 Gie\ss{}en,
	Germany
}

\emailAdd{philipp.isserstedt@physik.uni-giessen.de}
\emailAdd{christian.fischer@theo.physik.uni-giessen.de}
\emailAdd{thorsten.steinert@dwd.de}

\notes{%
	\note[$\dagger$]{%
		Present address:
		Frankfurt Consulting Engineers,
		Bessie-Coleman-Str.~7,
		60549 Frankfurt am Main,
		Germany.
	}
	\note[$\ddagger$]{%
		Present address:
		Deutscher Wetterdienst,
		Frankfurter Str.~135,
		63067 Offenbach am Main,
		Germany.
	}
}

\abstract{%
	In this contribution, we summarize a truncation-independent method to
	compute the equation of state within nonperturbative functional approaches.
	After demonstrating its viability, the method is applied to solutions
	obtained from a set of truncated Dyson--Schwinger equations for the quark
	and gluon propagators of ($2 + 1$)-flavor QCD to obtain thermodynamic
	quantities across the phase diagram of strong-interaction matter.
}

\FullConference{%
	FAIR next generation scientists -- 7th Edition Workshop (FAIRness2022)\\
	23--27 May 2022\\
	Paralia (Pieria, Greece)
}

\begin{document}

\maketitle

\section{\label{sec:introduction}%
	Introduction
}

Both the experimental and theoretical exploration of the phase diagram of
strong-interaction matter with its conjectured critical endpoint (CEP) is one
of the leading research goals of contemporary high-energy physics
\cite{Braun-Munzinger:2008szb,Friman:2011zz,Bzdak:2019pkr}. The equation of
state (EoS)---i.e., pressure, entropy density, and energy density---as a
function of temperature and chemical potential holds the desired information
to understand the various phases of QCD and is encoded in the thermodynamic
potential.

The phase structure of QCD at vanishing chemical potential is well understood
thanks to ab-initio lattice-QCD calculations \cite{Borsanyi:2013bia,%
HotQCD:2014kol}. At nonzero chemical potential, however, these are severely
hampered by the sign problem and complementary approaches are necessary. The
nonperturbative functional framework of Dyson--Schwinger equations (DSEs) is
well-suited for that, and significant progress has been achieved in recent
years; see, e.g., Refs.~\cite{Isserstedt:2019pgx,Gao:2020qsj,Gao:2020fbl,%
Isserstedt:2020qll,Gunkel:2021oya,Bernhardt:2021iql,Bernhardt:2022mnx} and
Ref.~\cite{Fischer:2018sdj} for a review.

Unfortunately, obtaining the EoS from DSEs is extremely difficult. Generally
speaking, the starting point for the derivation of every DSE is the first
derivative of the thermodynamic potential, and an integration is necessary to
get hold of the potential itself. Alas, this integration is only possible for
certain truncations. In this contribution, a truncation-independent method to
compute thermodynamic quantities from DSEs is summarized that we put forward
in Ref.~\cite{Isserstedt:2020qll}.

\section{\label{sec:method}%
	Thermodynamics from the quark condensate
}

At nonzero temperature $T$ and quark chemical potential $\mu$, all thermodynamic
information of QCD is encoded in its thermodynamic potential
\begin{equation}
	\label{eq:omega}
	\Omega(T, \mu)
	=
	-\frac{T}{V} \log \mathcal{Z}(T, \mu) \, ,
\end{equation}
where $V$ is the volume of the system and $\mathcal{Z}$ denotes the
grand-canonical partition function.%
\footnote{%
	For the sake of simplicity, here we consider only one (light) flavor.
}
The EoS follows from standard thermodynamic relations \cite{Kapusta:2006pm}.
For example, the pressure is given by
$P(T, \mu) = -(\Omega(T, \mu) - \Omega(0, 0))$
and the entropy density reads
$s(T, \mu) = \partial P(T, \mu) / \partial T$.

Though in principle fixed, the current quark mass $m$ can also be treated as an
external, variable quantity: $\Omega = \Omega(T, \mu; m)$. In the action, it
appears as the source for the quark-field bilinear $\overline{\psi} \psi$, and
the quark condensate is thus given by
$\qcond(T, \mu; m) = \partial \+ \Omega(T, \mu; m) / \partial m$.
This relation can be inverted, i.e., integrated with respect to the current
quark mass, and allows us to express the difference of the thermodynamic
potential evaluated at two arbitrary (unphysical) current quark masses
$m_1$, $m_2$ (w.l.o.g., $m_1 \?< m_2$) as an integral over the quark
condensate, viz.
\begin{equation}
	\label{eq:omega_condensate_relation}
	\Omega(T, \mu, m_2) - \Omega(T, \mu, m_1)
	=
	\int_{m_1}^{m_2} \dd m' \, \qcond(T, \mu; m') \, .
\end{equation}
This equation is of little practical use because both $\Omega$ and $\qcond$
are divergent. The divergence is contained in the vacuum contribution to
the thermodynamic potential, i.e., independent of temperature and chemical
potential. Thus, a derivative of Eq.~\eqref{eq:omega_condensate_relation}
with respect to $T$ yields a divergence-free equation that relates the
entropy density to the quark condensate according to
\begin{equation}
	\label{eq:entropy_condensate_relation_1}
	s(T, \mu; m_2) - s(T, \mu; m_1)
	=
	-\int_{m_1}^{m_2} \dd m' \,
	\frac{\partial \qcond}{\partial T}(T, \mu; m') \, .
\end{equation}

Finally, we set the lower integration limit to the physical current quark mass,
$m_1 = m$, and send the upper one to infinity, $m_2 \to \infty$. An infinitely
heavy quark becomes static and freezes out of the system, and the entropy
density is then simply the one of pure-gluonic Yang--Mills gauge theory,
$s(T, \mu; m_2 \to \infty) = \sYM(T)$, which is known to a high precision from
the lattice \cite{Borsanyi:2012ve}. We thus arrive at
\begin{equation}
	\label{eq:entropy_condensate_relation_2}
	s(T, \mu)
	=
	\sYM(T)
	+
	\int_m^\infty \dd m' \,
	\frac{\partial \qcond}{\partial T}(T, \mu; m') \, ,
\end{equation}
which establishes a general relation between the entropy density and the quark
condensate. We emphasize that Eq.~\eqref{eq:entropy_condensate_relation_2} is
obtained without any approximation and is therefore exact. Only the
temperature derivative of the quark condensate as a function of the current
quark mass (at fixed $T$ and $\mu$) is needed to compute the entropy density.
This renders Eq.~\eqref{eq:entropy_condensate_relation_2} quite general and
not restricted to a specific approach, though it is particularly useful within
the DSE framework where computing the thermodynamic potential is extremely
difficult, while obtaining $\qcond(T, \mu; m)$ is straightforward.

\section{\label{sec:dse}%
	Dyson--Schwinger equations
}

The quark condensate---sole input of
Eq.~\eqref{eq:entropy_condensate_relation_2}---is obtained from the
nonperturbative quark propagator $S(T, \mu; m)$ through
$\qcond(T, \mu; m) = \Tr S(T, \mu; m)$, where the trace is understood in the
functional sense over flavor, color, Dirac, and momentum space d.o.f. We obtain
$S$ by solving a set of truncated DSEs that takes the backcoupling of quarks
onto the gluon explicitly into account, which allows for a consistent mass and
flavor dependence of all results. Equally important, the gluon becomes
sensitive to the chiral dynamics of the quarks. This system has been studied
extensively in other works (see, e.g., Refs.~\cite{Isserstedt:2019pgx,%
Bernhardt:2021iql,Fischer:2018sdj} and references therein) and predicts a
second-order CEP at temperature $\TCEP \approx \SI{119}{\MeV}$ and baryon
chemical potential $\muBCEP \approx \SI{495}{\MeV}$.

\begin{figure}[t]
	\centering%
	\includegraphics[scale=0.92,trim=0 -5.8mm 0 0]{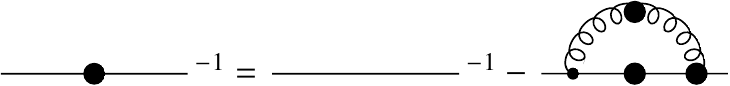}%
	\hfill%
	\includegraphics[scale=0.92]{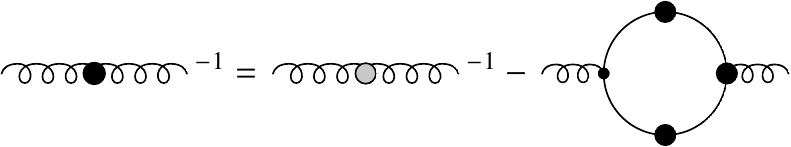}%
	\caption{\label{fig:dse}%
		DSEs for the quark (left) and gluon (right) propagators. The former
		(latter) are denoted by solid (curly) lines while a large black dot
		indicates nonperturbative quantities. In the gluon DSE, the gray dot
		represents all diagrams with no explicit quark content. Feynman
		diagrams were drawn with \textsc{JaxoDraw} \cite{Binosi:2008ig}.
	}
\end{figure}

In more detail, the nonperturbative quark and gluon propagators each obey a
DSE that is shown diagrammatically in Fig.~\ref{fig:dse}. Both contain 
higher-order correlation functions, e.g., the nonperturbative quark-gluon
vertex, which satisfy their own DSEs. To cope with this infinite tower of
coupled equations, we truncate as follows: (i) in the gluon DSE, all diagrams
with no explicit quark content are replaced by fits to quenched,
temperature-dependent lattice results; (ii) for the nonperturbative quark-gluon
vertex, an ansatz is employed whose infrared dynamics are guided by a
Slavnov--Taylor identity for the full vertex while its behavior in the
ultraviolet is fixed by demanding the correct running of the propagators at
large perturbative momentum scales. The quark-loop diagram is evaluated
explicitly, thereby unquenching the gluon.

This system is solved numerically for $2 + 1$ flavors, which yields the
nonperturbative quark and unquenched gluon propagators at arbitrary $T$ and
$\muB$. For the sake of brevity, we refrain from showing explicit expressions
and refer the reader to Refs.~\cite{Isserstedt:2019pgx,Fischer:2018sdj}
for details.

\section{\label{sec:results}%
	Results and discussion
}

\begin{figure}[t]
	\centering%
	\includegraphics[scale=1.0]{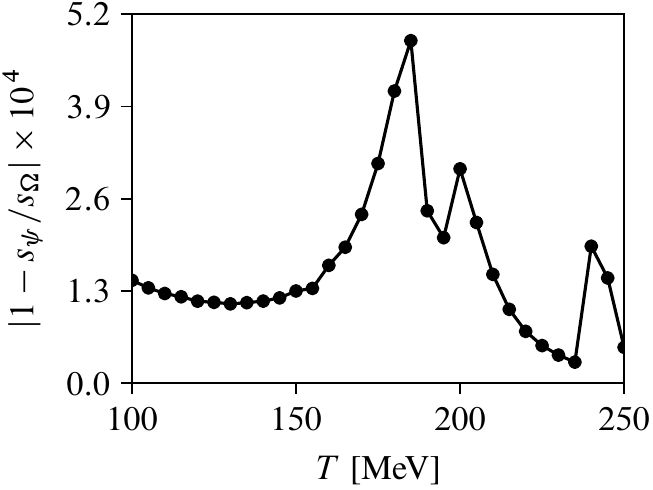}%
	\hspace{0.5em}%
	\includegraphics[scale=1.0]{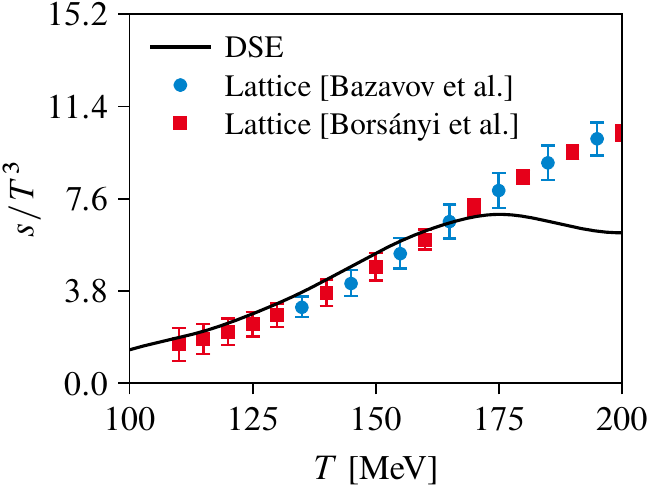}%
	\\[0.5em]%
	\includegraphics[scale=1.0]{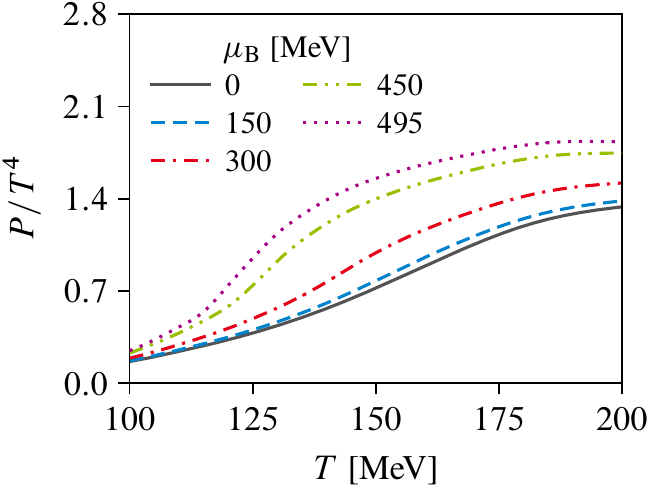}%
	\hspace{0.5em}%
	\includegraphics[scale=1.0]{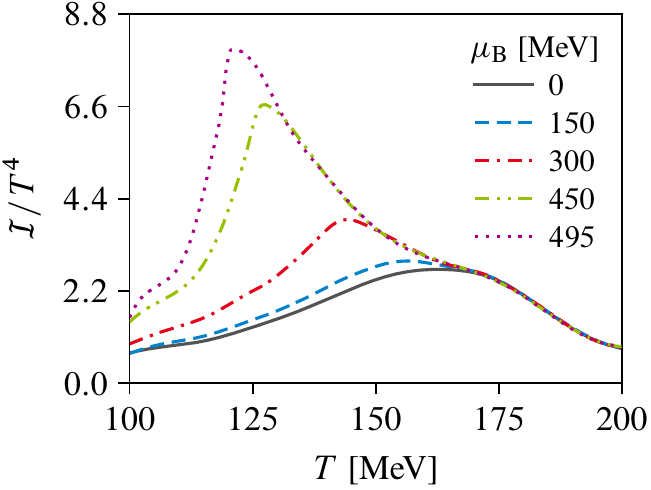}%
	\caption{\label{fig:results}%
		\textit{Upper left:}
		relative error between the NJL entropy density obtained from the
		quark condensate through Eq.~\eqref{eq:entropy_condensate_relation_2}
		($s_\psi$) and directly from the thermodynamic potential ($s_\Omega$).
		\textit{Upper right:}
		entropy density at $\muB = 0$ in comparison to lattice results
		\cite{Borsanyi:2013bia,HotQCD:2014kol}.
		\textit{Lower left and right:}
		pressure and interaction measure, respectively, at different chemical
		potentials up to the CEP.
		All diagrams are adapted from Ref.~\cite{Isserstedt:2020qll}.
	}%
\end{figure}

\paragraph*{Proof of principle}

In order to demonstrate the viability of the method described in
Sec.~\ref{sec:method}, we use a two-flavor Nambu--Jona-Lasinio (NJL) model in
mean-field approximation \cite{Hatsuda:1994pi}. It has the advantage that the
thermodynamic potential can be computed explicitly and is given in a closed
form. We are therefore in a position to compare the NJL entropy density obtained
directly from the thermodynamic potential with the one resulting from
Eq.~\eqref{eq:entropy_condensate_relation_2}.%
\footnote{%
	Here, $\sYM = 0$ because gluons are no active d.o.f.\ in the NJL model.
}

We find that both results cannot be distinguished by the eye, and the relative
error between them is shown in the upper left diagram of 
Fig.~\ref{fig:results}: it is smaller than $0.05\%$ across the whole temperature
range. This shows that our method is reliable and works well.

\paragraph*{EoS from DSEs}

Now, we summarize our thermodynamic results, which are discussed in detail in
Ref.~\cite{Isserstedt:2020qll}, obtained using the method described in
Sec.~\ref{sec:method} with quark-condensate data extracted from the DSE
framework of Sec.~\ref{sec:dse}.

Our result for the entropy density---the direct outcome of our method---at
vanishing chemical potential is shown in the upper right diagram of
Fig.~\ref{fig:results} together with continuum-extrapolated lattice results
\cite{HotQCD:2014kol,Borsanyi:2013bia}. For $T \lesssim \SI{170}{\MeV}$, it is
monotonically increasing and, though with a slight overshoot, in satisfying
agreement with the lattice data. At higher temperatures, we observe an
unphysical, nonmonotonous behavior. The reason for that can be traced back to
our vertex ansatz: it overestimates the strength of the quark-gluon
interaction at high temperatures and/or chemical potentials, which should
become continuously weaker due to thermal screening. Addressing that issue
is not the aim of this work, neither the delivery of a high-quality EoS;
though we would like to note that our setup yields satisfying results in the
temperature range $\numrange{100}{160} \, \si{\MeV}$, i.e., below and around
the pseudocritical chiral transition temperature. More important, the key
message is that the method of Sec.~\ref{sec:method} allows us to obtain the
entropy density from an elaborate DSE framework that does not admit an
explicit calculation of the thermodynamic potential.

Having the entropy density $s(T, 0)$ at vanishing chemical potential at hand,
the pressure follows thermodynamically consistent from
\begin{equation}
	\label{eq:pressure}
	P(T, \muB)
	=
	P(T_0, 0)
	+
	\int_{T_0}^T \dd T' \+ s(T', 0)
	+
	\int_0^{\muB} \dd \mu' \+ n(T, \mu') \, ,
\end{equation}
where the baryon number density $n(T, \muB)$ is calculated using the framework
of Ref.~\cite{Isserstedt:2019pgx}, and $P(T_0, 0) / \+ T_0^{\+ 4} = 0.242$ at
$T_0 = \SI{110}{\MeV}$ \cite{Borsanyi:2013bia}. Our results for the pressure
at different chemical potentials ranging from zero to the CEP value
$\muBCEP = \SI{495}{\MeV}$ as functions of $T$ are depicted in the lower left
diagram of Fig.~\ref{fig:results}. The pressure gets larger with increasing
chemical potential across the whole temperature range, although the increase is
less noticeable at low temperatures. Its inflection point with temperature can
be used to define the pseudocritical chiral transition temperature in the
crossover region of the phase diagram and is compatible with other definitions
like the peak position of the chiral susceptibility or the inflection point
of the quark condensate. Approaching the CEP, a kink develops around the
corresponding (chemical-potential dependent) pseudocritical chiral transition
temperature, which is most pronounced close and at the CEP. Beyond that point,
the pressure rises with a steep slope compared to low chemical potentials.

Since the entropy density is used to obtain the pressure, the latter inherits
the erroneous high-temperature behavior from the former. This manifests in the
fact that our results for the pressure saturate at too low temperatures and
(well) below the Stefan--Boltzmann limit. However, the very same key message
as stated above in case of the entropy density applies here.

Finally, our results for the interaction measure
$\mathcal{I} = \varepsilon - 3 P$ are shown in the lower right diagram of
Fig.~\ref{fig:results}, where $\varepsilon$ denotes the energy density, which
is obtained by a Legendre transform of the pressure. At small chemical
potentials, the interaction measure is shape consistent with lattice results
and experiences as a function of temperature a strong increase from
intermediate $\muB$ toward $\muBCEP$. Close to and at the CEP, the slope
becomes (near-)infinite at the corresponding critical temperature, and the
peaklike structure with a large magnitude indicates that nonperturbative
effects are manifest in this region of the QCD phase diagram.

\paragraph*{Closing remarks}

We summarized a truncation-independent method to compute the EoS within
nonperturbative functional approaches. At its heart lies an exact relation
between the entropy density and the quark condensate. The method is
particularly useful in the context of DSEs because it allows for the
calculation of thermodynamic quantities within truncations that do not admit
an explicit calculation of the thermodynamic potential---all current
state-of-the-art DSE calculations \cite{Isserstedt:2019pgx,Gao:2020qsj,%
Gao:2020fbl,Gunkel:2021oya} are of that kind and future ones will certainly
be like that, too.

\acknowledgments

This work was supported by HGS-HIRe for FAIR, the GSI Helmholtzzentrum
f\"{u}r Schwerionenforschung, and the BMBF under contract 05P18RGFCA.
We acknowledge computational resources provided by the HPC Core Facility
and the HRZ of the Justus-Liebig-Universit\"{a}t Gie\ss{}en.

\bibliographystyle{JHEP}
\begin{small}
\bibliography{Isserstedt_Fairness}
\end{small}

\end{document}